# Poset representation and similarity comparisons of systems in IR

Christine MICHEL

Centre d'Étude des Médias
MSHA - Université Bx3
10, Esplanade des Antilles
33607 PESSAC Cedex
FRANCE
Tel : 33 5 56 84 68 13

Christine.Michel@montaigne.u-bordeaux.fr

**Abstract.** In this paper we are using the poset representation to describe the complex answers given by IR systems after a clustering and ranking processes. The answers considered may be given by cartographical representations or by thematic sub-lists of documents. The poset representation, with the graph theory and the relational representation opens many perspectives in the definition of new similarity measures capable of taking into account both the clustering and ranking processes. We present a general method for constructing new similarity measures and give several examples. These measures can be used for semi-ordered partitions; moreover, in the comparison of two sets of answers, the corresponding similarity indicator is an increasing function of the ranks of presentation of common answers.

## 1. Introduction

Informational objects are support or mediator used to improve perception of information for people working with computer. These objects are simple texts, combined or not with fixed or animated pictures. They are produced either manually by one or several authors, or dynamically by information retrieval systems (i.e. like answers produced by an IR system). These objects may be viewed as aggregates of indexed elements, generally called fragments, stocked in databases. The strategies for constructing such objects may vary with the system so these objects may have very different structures; however one can isolate formally three main characteristics; the degree of aggregation (clustering process), the scheduling (ranking process) and the repetition of fragments. Very often, developers and ergonomists use these three characteristics to improve the usefulness, the relevance and the efficiency of these systems. Paradoxically, lacking some appropriate formalism, these characteristics are rarely taken into account in the evaluation or comparison of sets of informational objects constructed by two different systems. Our aim is to give a conceptual representation of these characteristics in order to produce numerical indicators that are capable to quantify the degree of similarity between two sets. We have chosen to use partially ordered set (poset) representations and graph theory. Indeed, *"it is possible to model the logical structure of any real-world systems using concept developed in the branch of mathematics known as graph theory"* (Furner, Ellis and Willett, 1996). With this representation, we propose new similarity measures that take into account the clustering and ranking processes present in the IR systems. Theses measures may be useful for large-scale quantitative evaluation experiments.

## 2. Chain, antichain, partition and poset in IR

We begin with some definitions taken from (Davey and Priestley, 2002) and (Schreider, 1975). Let P be a set. An *order* (or *partial order*) on *P* is a binary relation ≤ which is reflexive, antisymetric and transitive. A set *P* equipped with an ordered relation is written $\langle P, \leq \rangle$ and is called an *ordered set* (or a *partially ordered set*). Some authors use the shorthand *poset*. On any set, = is an order, the *discrete* order. The relation < constructed from $\leq$ and defined by $x < y$ *if and only if* $x \leq y$ *and* $x \leq y$, is called a strict inequality. $\langle P, < \rangle$ is called a *strictly ordered set*. $\langle P, \leq \rangle$ is called a *chain* if any two element of P are comparable. Alternatives names for a chain are *linearly ordered set* or *totally ordered sets*. Conversely, $\langle P, \leq \rangle$ is an *anti-chain* if $x \leq y$ in P only if $x = y$, so any $\langle P, = \rangle$ where any two element are comparable is an *antichain*.

We shall call a collection $\{P_1, P_2, ... P_u\}$ of non empty subsets of P, a *partition* if $P = \bigcup_{i=1}^{u} P_i$ and $\forall i, j \in [1, u], i \neq j \quad P_i \cap P_j = \emptyset$

The subsets $P_1, P_2, ... P_u$ are called the *classes* of the given partition. A relation $A$ in a set $P$ is called an *equivalence relation* if there exists a partition $\{P_1, P_2, ... P_u\}$ of the set $P$, such that the relation $xAy$ holds if and only if $x$ and $y$ belong to some common class $P_i$ of that partition.

The simplest answers given by information retrieval systems is an antichain or a strictly ordered chain of elements (documents, web sites, …), as it is shown in figure 1 and 2 where points represent the elements presented to the users by the system.

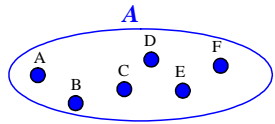

**Figure 1.** Antichain

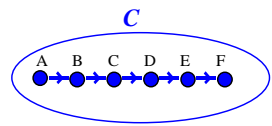

**Figure 2.** Strictly ordered Chain

The search for new types of presentation and new interfaces revealed other forms of presentation, like sub lists of themes or cartographic clusters, characterized by regroupings the elements according to a semantic unit. "*Cluster analysis is a technique that allow the identification of groups or clusters of similar objects in multi-dimensional space. It was initially introduced in the field of Information Retrieval as a mean of improving the efficiency of serial search*" (Tombros and Van Rijsbergen, 2001). In information retrieval, the elements are presented only one time, so sets produced are partitions where the clusters are the classes (see figure 3).

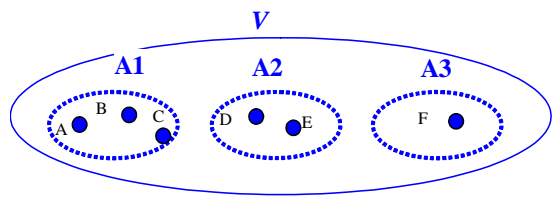

**Figure 3.** Partition

This characteristic is specific to the IR process and databases area. In a considerable number of other contexts, the informational objects are built from elements which can be repeated: a text is composed of words which can be repeated, a dynamic Web page is composed of elements, image or hypertext link which, for ergonomics reasons, can be repeated. We will not treat these cases and restrict ourselves to the context of partition.

Usually, systems use a combination of ranking and clustering processes but the partition proposed as answers is rarely given by totally ordered sets where the classes are ordered and are themselves chain: in general, sets of answers have the structure of a poset. Two cases can arise:

- Classes themselves are ordered and can be compared but the elements inside the classes are not ordered (figure 4); we will speak of *ordered partition* (Lebanon and Lafferty, 2002).

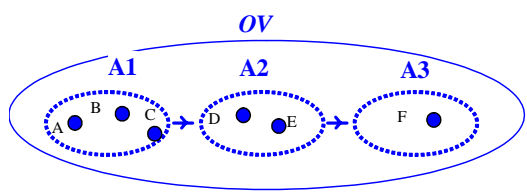

**Figure 4.** Ordered partition

It is typically the case that arises with the informational system Spirit (Fluhr, 1997) or the web search engine Vivisimo (http://vivisimo.com). Elements are gathered in classes according to whether or not they contain a combination of the informational words given in the question; elements have the same importance in the class because they all contain the same combination of words; the classes have more or less importance depending on the number of informational words, characterizing them.

- Classes are not ordered but the elements inside a given class are (figure 5), we will speak of *partition of chain*.

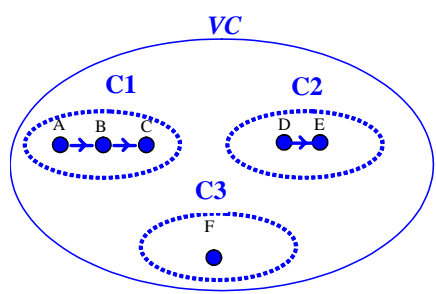

**Figure 5.** Partition of chain

It is typically the case that arise in cartographic system using semantic network, like Kartoo (http://kartoo.com/) or Webbrain (http://www.webbrain.com), or other meta criteria, like the type of leading support (summarized, vulgarization article) or the source of production (commercial Web site, personal pages, magazine...) (Zamir and Etzioni, 1999) : classes are formed by regrouping documents according to their degree of similarity; documents are then ordered in the classes, the most representative of the class being presented the first.

For large-scale quantitative evaluation processes it is necessary to compare system's answers given by chain, antichain, partition or poset.

How to do it and what are the criterion to take into account in order to determine the degree of similarity between them?

## 3.     Similarity comparison of sets

*"The concept of similarity and similarity relations play a fundamental role in many fields of pure and applied science. The notion of a metric or distance d(x,y) between object x and y has long been used in many context as a measure of similarity or dissimilarity between element of a set."* (Finnie and Sun, 2002)

Ellis et al. (Ellis, Furner-Hines and Willett, 1994) have considered that there is two types of similarity formulas: the first ones are based on contingency-table notation and the second ones are based on set-theoretic notation. Ellis give another taxonomy of measures and group them into three classes: the values of *distance coefficients* vary inverse proportionally with the degree of similarity: some examples are the mean difference sum, the mean Euclidean (called also average distance). The values of *association coefficients* vary in direct proportionally with the degree of similarity: some examples are the Jaccard index, the dice index or the Cosine. The values of *correlation coefficients* generally vary from +1 (indicating that any change in the attributes of one object is be accompanied by an identical change in the attributes of the other) to –1 (indicating that any change in one is accompanied by an equal and opposite change in the other), one example is the coefficient of Pearson.

Our own taxonomy is based on the type of sets to which the similarity comparison can be applied (i.e. non ordered set, totally ordered sets, partially ordered sets) because the criterion of similarity are not the same in each case. When the system uses no order and no clustering techniques, the degree of similarity of the answers is grows with the number of common documents. When the system uses clustering techniques, answers are partitions and the similarity of two answers grows in function of the number of pair of documents found within the same partitions, in both sets. When the systems propose as an answer, a totally ranked list of documents, the similarity has to grow in function of the number of common documents found with the highest ranks. And finally, when the systems propose a partially ordered answer, the similarity criterion must take into account both the similarity between the partitions and the similarity between the ranks (by giving more weight to common documents found with higher rank). We present now some similarity measures for each of these cases.

Remark : In the following we will use the convention: A is for an antichain, C for a chain, V for a partition, OV for an ordered partition and VC for a partition of chain.

### 3.1     Antichain Case

The similarity between two antichains, i.e. sets without order and without partition, is usually based on the number of elements that they can have in common, calculated, if A1 and A2 are the sets to be compared, by the cardinal of $A1 \cap A2$ (noted $|A1 \cap A2|$ ). The similarity is quantified by a number ranging between 0 and 1, 0 means that the compared sets do not have any common elements i.e. that $A1 \cap A2 = \emptyset$, and 1 means classically that they are strictly identical i.e. $A1 = A2$. In fact, this last natural condition may be false when specific roles are given to A1 and A2 : in the case of the recall and the precision for instance. Indeed, the recall (R) is the proportion of relevant documents found compared to the number of relevant documents. Let us consider that A1 is the set of documents found and A2 the set of the relevant documents. R=1 means that all the relevant documents have been found i.e. that $A2 \subset A1$ and not that *only* the relevant documents have been found i.e. that $A2 = A1$. In the same manner, P=1 means that all the documents found are relevant i.e. $A1 \subset A2$ and not that *only* the documents found are relevant. This observation makes it possible to separate similarity measures in two groups: *strong and weak measures* formally defined in (Egghe and Michel, 2002). Very simply, strong measurements are characterized by a strict identity of the compared sets if the proximity is 1, weak measurements are defined by an inclusion of the sets in similar case. The main strong similarity measures are: the Jaccard's coefficient, the Dice's coefficient, the Cosine, the measures of efficiency built as a combination or an average of recall and precision (the most general is the Generalized Dice's index (Van

Rijsbergen, 1981).). The main weak similarity measures are the recall and the precision. All these measures are defined in (Boyce, Meadow and Kraft, 1994).

### 3.2 The Case of Partitions

In the case of sets with partition, the most usual indicator used to measure the similarity between them (Saporta and Youness, 2002) is the Rand index: it represents the global percentage of pair in accord; i.e. witch are on the same class in each partition.

*3.2.1 The Rand index*

Indeed, let V1 and V2 be two partitions of n objects where p is the number of classes of V1, and q the number of classes of V2. From each pair of objects the four types of situations can arise:
- Type 1: pairs belonging to the same class of V1 and to the same class of V2
- Type 2: pairs belonging to different classes of V1 but to the same class of V2
- Type 3: pairs belonging to the same class of V1 but to different classes of V2
- Type 4: pairs belonging to different classes of V1 and to different classes of V2.

If the respective frequencies of these four cases are a, b, c, d we have: $a+b+c+d = \frac{n(n-1)}{2}$. We note also $A = a+d$ the number of agreements and $D = b+c$ the number of disagreements.
The Rand index $R_{Rand}$ is :

$$R_{Rand} = \frac{2A}{n(n-1)} \quad (1)$$

$R_{Rand} = 1$ if the partitions are really closed, $R_{Rand} = 0$ else.

*3.2.2 Asymmetric Rand index*

If the sets don't have the same number of partitions, we can use the asymmetrical Rand index (Chavent, Lacomblez and Patrouille, 2001) (noted $R_{RandAsy}$) to define the thinnest partition. If V1 has more classes than V2, V1 is said to be thinnest than V2 if V1 is included into V2, which means $\forall u = 1,..., p, \exists v$ where $V1_u \subseteq V2_v$

In this case, if $\forall u = 1,..., p, \exists v$ where $V1 \subseteq V2$ then $R_{RandAsy} = 1$.
The computation of the asymmetrical Rand index is :

$$R_{RandAsy}(V1,V2) = \frac{A+b}{n(n-1)} \quad (2)$$

*3.2.2.1 Jaccard index*

The jaccard index can be written in the case of partitioning sets comparison. It is given by :

$$J(V1,V2) = \frac{a}{a+b+c} \quad (3)$$

## 3.3 Strict Chain's Case

The arrival of IR systems ordering completely or partially the documents according to a value of relevance made the construction of similarity measures reflecting this parameter necessary. Experiments in this case usually consider only the set constituted by the first n documents, and compute usual measures (without the order's information) on this new antichain. It is the solution chosen in the case of the evaluation protocol's TREC. Some examples of measures used in it (Buckley and Voorhees, 2000) are:

*Prec(~):* Precision at cut-off level A, for A = 1, 2, 5, 10, 15, 20, 30, 50, 100, 300, 1000. A cut-off level is a rank that defines the retrieved set; for example, a cut-off level of ten defines the retrieved set as the top ten documents in the ranked list.

*Recall(1000):* Recall after 1000 documents have been retrieved.

*Prec at .5 Recall:* Precision after half the relevant document have been retrieved.

*R-Prec:* Precision after R documents has been retrieved where R is the number of relevant documents for the current topic.

*Average Precision:* The mean of the precision scores obtained after each relevant document is retrieved, using zero as the precision for relevant documents that are not retrieved.

One will find more than 80 measurements of the same type and their comparative analysis in (Voorhees and Harman, 1998).

Another way to compare chain is to compute the coefficient of correlation of row of Spearman $R_s$ - representative of the number of permutations to be carried out to replace the common elements of two sets in the same order - and multiply it by a classical similarity indicator (Jaccard, Cosine, Dice, …) called D where sets to be compared are considered without ranks. An experimentation of this combination is presented in (Tague-Sutcliffe, 1995) and can be written:

$$D_O(C1, C2) = R_s(C1, C2) \times D(C1, C2) \quad (4);$$

here C1 and C2 are the two chains to be compared; they are ordered elements of $E = \{x_1, ..., x_i, ..., x_n\}$.

$R_s$ is written :

$$R_s(C1, C2) = 1 - 6 \frac{\sum_{i=1}^{n}(|x_i^1 - x_i^2|)^2}{n(n^2 - 1)} \quad (5)$$

where $x_i^1$ is the rank of the item $x_i$ in the set C1 and $x_i^2$ is the rank of the item $x_i$ in the set C2.

The correlation of row $R_s$ can be replaced by the Kendall tau $\tau$.

$$\tau = \frac{2S}{n^2 - n} \quad (6)$$

Where S is the sum of the scores given to each pairs of item compared. A score of +1 is given to a pair if they are on the same order in C1 and C2; a score of -1 is given else.

These two types of construction, with the cut-off level and the combination of $R_s$ or $\tau$ and D are not precise enough: the first only takes into account a fraction of the answer; the second does not distinguish between the finding of common elements in the first ranks and in the last. Moreover, these two types of measures are not applicable for posets.

## 3.4 Poset's Case

The formalism of poset has been used in IR by (Sparck Jones, Walker and Robertson, 1998) and (Jourliny, Johnsonz , Sparck Jones and Woodland, 1999) to improve the research of information in the case of multimedia document retrieval. The structure and the property of poset have been used in many domain such as mathematical psychology, social sciences, decision making, expert judgment, chemistry, biology information processing, retrieval, artificial intelligence e.g. (knowledge structures), and in studies about preference relation and aggregation of preferences. Many papers presented work with the similarity and dissimilarity of structures defined by the poset representation but we haven't found any paper in the context of evaluation of IR system, in which the similarity criterion for two posets takes into account the fact that the similarity is greater if common elements are found in the higher ranks of the posets than in the smaller. We have seen (Michel, 2000) that this point constitutes an important criterion in the case of IR systems evaluation.

In the case of ordered partition posets, Egghe et al. (Egghe and Michel, 2003) propose a construction of weak and strong ordered similarity measures taking this criterion into account. The construction uses the formalism of fuzzy sets (Zadeh, 1979). The principle is to rewrite the definition of the ordered partition by the membership function of the fuzzy set logic defined below.

Let $OV = (A_i)_{i \in N}$ be an ordered partition. We consider $U_{OV} = \bigcup_{i=1}^{n} A_i$ a poset equipped with the membership function

$$P_{U_{ov}} = \varphi(i) \Leftrightarrow x \in A_i \text{ where } \varphi \text{ is strictly decreasing with i.} \tag{7}$$

By using this definition it is possible to rewrite classical intersection and union and then also classical similarity measures.

For example, if $OV1 = (A1_i)_{i \in N}$ and $OV2 = (A2_i)_{i \in N}$, by using the membership function $\varphi(i) = \frac{1}{2^{i-1}}$, we have (demonstrations are in (Egghe and Michel, 2003)):

$$|U_{OV1} \cap U_{OV2}| = \sum_{i=1}^{\infty} \sum_{j=1}^{\infty} |A1_i \cap A2_j| \frac{1}{2^{\max(i,j)-1}} \tag{8}$$

$$|U_{OV1}| = \sum_{i=1}^{\infty} |A1_i| \frac{1}{2^{i-1}} \tag{9}$$

$$|U_{OV2}| = \sum_{j=1}^{\infty} |A2_j| \frac{1}{2^{j-1}} \tag{10}$$

$$|U_{OV1} \cup U_{OV2}| = \sum_{i=1}^{\infty}\sum_{j=1}^{\infty} |A1_i \cap A2_j| \frac{1}{2^{i-1}} + \sum_{i=1}^{\infty}\sum_{j=1}^{j-1} |A1_i \cap A2_j| \frac{1}{2^{j-1}} + \sum_{i=1}^{\infty} \left| A1_i \setminus \bigcup_{j=1}^{\infty} A2_j \right| \frac{1}{2^{i-1}} + \sum_{j=1}^{\infty} \left| A2_j \setminus \bigcup_{i=1}^{\infty} A1_i \right| \frac{1}{2^{j-1}} \tag{11}$$

And so it is possible to write new ordered similarity measures by using these definitions of intersection, cardinality and unions. Theses new measures may be constructed from the classical weak and strong similarity measures. For example, the ordered coefficient of Jaccard will be written as:

**Ordered Jaccard:**

$$J_F(OV1, OV2) = \frac{|U_{OV1} \cap U_{OV2}|}{|U_{OV1} \cup U_{OV2}|} = \frac{\sum_{i=1}^{\infty}\sum_{j=1}^{\infty} |A1_i \cap A2_j| \frac{1}{2^{\max(i,j)-1}}}{\alpha} \tag{12}$$

with $\alpha = \sum_{i=1}^{\infty} \sum_{j=1}^{\infty} |A1_i \cap A2_j| \frac{1}{2^{i-1}} + \sum_{i=1}^{\infty} \sum_{j=1}^{j-1} |A1_i \cap A2_j| \frac{1}{2^{j-1}} + \sum_{i=1}^{\infty} |A1_i \setminus \bigcup_{j=1}^{\infty} A2_j| \frac{1}{2^{i-1}} + \sum_{j=1}^{\infty} |A2_j \setminus \bigcup_{i=1}^{\infty} A1_i| \frac{1}{2^{j-1}}$ (13)

The reader will find in (Egghe and Michel, 2003) formulas of many other similarity measures like; the Ordered Dice, the Ordered Cosines, the Ordered N measure, the Ordered overlap 2, the Ordered overlap 1, the Ordered recall, the Ordered precision. Let us recall that they can't be used in the case of partition of chain.

## 4. Graph theory and relational representations

*"One of the most useful and attractive features of ordered sets is that, in the finite case at least, they can be 'drawn'"* (Davey and Priestley, 2002). Indeed, graph theory allows us to draw the poset representations. In the IR context graph theory is often used to represent hypertext structures (Ellis, Fumer-Hines and Willett, 1994b).

A direct graph G can be defined by an ordered pair of the form $G = <V, E>$, where V is an object set given by $V = \{v_1, v_2, ... v_p\}$, each of whose $p$ members is called vertex, and E is a family of two-member subset of V given by $E = \{\{v_i, v_j\}_1, \{v_i, v_j\}_2, ..., \{v_i, v_j\}_r\}$, each of the r member is called a direct edge. An edge "point" from the first coordinate of the pair to the second coordinate and represent a relation between the two objects (Furner, Ellis, and Willett, 1996). If the relation is noted R we will write $v_i R v_j$ if there is a relation between $v_i$ and $v_j$.

Let us considered the set $V = \{A, B, C, D, E, F\}$ and the three following relations: $R_\in$ =*'is on the same cluster than'*, $R_>$ =*'is greater than'* and $R_{rank}$ =*has the same rank than'*.

In the graph representation the relation $R_\in$ will be represented by plain arrow ⟶ , $R_\geq$ by dished arrows ---> and $R_{rank}$ by arrows with points ·······>.

Note: In the following graphs, loops are not represented in order to let the graph readable.

The set V in figure 3 is representing by the graph $G1 = <V, R_\in>$ in figure 6.

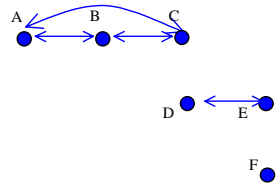

**Figure 6.** G1 Graph of V with $R_\in$

The set C in figure 2 is representing by the graph $G2 = <V, R_>>$ in figure 7.

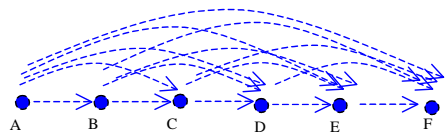

**Figure 7.** G2 Graph of C with $R_>$

The poset create some problems because its construction is made both by a clustering process and by a ranking process; so the set may be partially expressed in terms of $R_\in$, $R_>$ or $R_{rank}$. Indeed, the ordered partition *OV* given in figure 4 may be represented by the 2 relations $R_\in$ and $R_>$ like in the following figure.

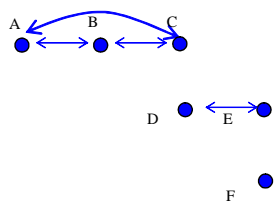

**Figure 8.** G3 Graph of OV with $R_\in$

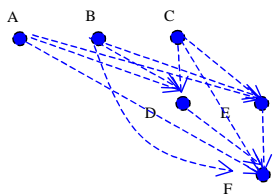

**Figure 9.** G4 Graph of OV with $R_>$

We have not drawn the graph with $R_{rank}$ because it is similar to the one with $R_\in$.

The graph with the tow relations is :

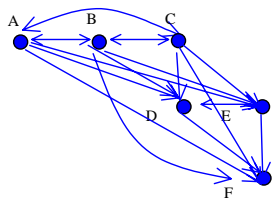

**Figure 10.** G5 Graph of OV with $R_\in$ and $R_>$

We can remark that $G5 = G3 \oplus G4$ where $\oplus$ is the linear sum defined in (Davey and Priestley, 2002). G5 represents the relation $R_\geq$ =*'is greater than equal to,* witch is the addition of $R_>$ and $R_\in$ (Schreider, 1975) : $R_\geq = R_> + R_\in$.

Let us considered now the partition of chain *VC* shown in figure 5. The information on it is represented by the 3 relations $R_\in$, $R_>$ and $R_{rank}$ as in the following figures.

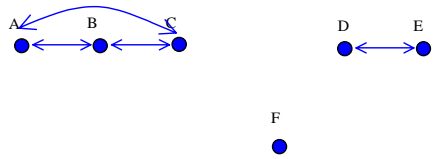

**Figure 11.** Graph of VC with $R_\in$

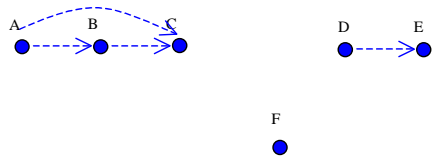

**Figure 12.** Graph of VC with $R_>$

We can remark that $R_> \subset R_\in$ .

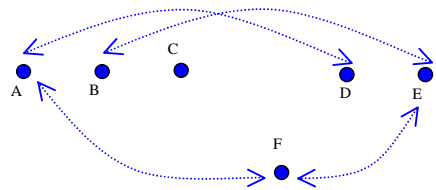

**Figure 13.** Graph of VC with $R_{rank}$

With the combination of two relations, the information in the set VC can be drawn as figure 14 or figure 15.

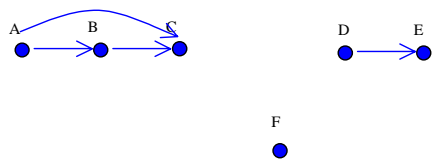

**Figure 14.** Graph of VC with $R_\in$ and $R_>$

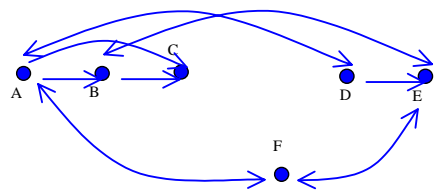

**Figure 15.** Graph of VC with $R_>$ and $R_{rank}$

We can remark that, with the aggregation relation $R_{rank}$ and the ordered relation $R_>$, *VC* can be viewed as an ordered partition:

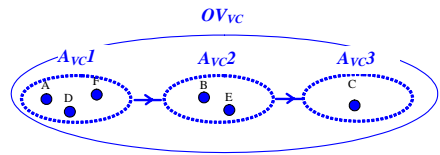

**Figure 16.** Representation of a partition of chain as an ordered partition with the aggregation's relation $R_{rank}$

*4.1.1 Numerical representation*

*"In order that a graph may be analyzed by computer it is necessary to represent it on a machine-readable format, which normally involves some form numerical matrix. For any direct graph we can define an adjacency matrix **A** consisting of $p \times p$ elements of the forms $a_{ij}$. Each of these elements has a value of 1 (if the set of links contains a member pointing from $v_i$ to $v_j$), 0 (if the set of links does not contain a member pointing from $v_i$ to $v_j$) or null (if i=j, i.e. if the node $v_i$ is the same as the node $v_j$)"* (Ellis, Fumer-Hines and Willett, 1994b). In his definition Ellis considers that the loops to have the value "null". We will prefer, in our case, considered that they give the value "1" in order to have pure binary matrix.

The adjacency matrix is usually noted $\mathbf{A}_G$ (Umeyama, 1988). In the case of the graph $G = <V, E>$, $\mathbf{A}_G$ is a $p \times p$ matrix defined as follows:

$$\mathbf{A}_G = [a_{ij}] \quad \begin{cases} a_{ij} = 1 & if \ \{v_i, v_j\} \in E \\ a_{ij} = 0 & else \end{cases}$$

The relational representation in table of the preceding examples are :

**Table 1.** Relational representation of figure 6

|   | A | B | C | D | E | F |
|---|---|---|---|---|---|---|
| A | 1 | 1 | 1 | 0 | 0 | 0 |
| B | 1 | 1 | 1 | 0 | 0 | 0 |
| C | 1 | 1 | 1 | 0 | 0 | 0 |
| D | 0 | 0 | 0 | 1 | 1 | 0 |
| E | 0 | 0 | 0 | 1 | 1 | 0 |
| F | 0 | 0 | 0 | 0 | 0 | 1 |

**Table 2.** Relational representation of figure 7

|   | A | B | C | D | E | F |
|---|---|---|---|---|---|---|
| A | 1 | 1 | 1 | 1 | 1 | 1 |
| B | 0 | 1 | 1 | 1 | 1 | 1 |
| C | 0 | 0 | 1 | 1 | 1 | 1 |
| D | 0 | 0 | 0 | 1 | 1 | 1 |
| E | 0 | 0 | 0 | 0 | 1 | 1 |
| F | 0 | 0 | 0 | 0 | 0 | 1 |

**Table 3.** Relational representation of figure 10

|   | A | B | C | D | E | F |
|---|---|---|---|---|---|---|
| A | 1 | 1 | 1 | 1 | 1 | 1 |
| B | 1 | 1 | 1 | 1 | 1 | 1 |
| C | 1 | 1 | 1 | 1 | 1 | 1 |
| D | 0 | 0 | 0 | 1 | 1 | 1 |
| E | 0 | 0 | 0 | 1 | 1 | 1 |
| F | 0 | 0 | 0 | 0 | 0 | 1 |

**Table 4.** Relational representation of figure 14

|   | A | B | C | D | E | F |
|---|---|---|---|---|---|---|
| A | 1 | 1 | 1 | 0 | 0 | 0 |
| B | 0 | 1 | 1 | 0 | 0 | 0 |
| C | 0 | 0 | 1 | 0 | 0 | 0 |
| D | 0 | 0 | 0 | 1 | 1 | 0 |
| E | 0 | 0 | 0 | 0 | 1 | 0 |
| F | 0 | 0 | 0 | 0 | 0 | 0 |

**Table 5.** Relational representation of figure 15

|   | A | B | C | D | E | F |
|---|---|---|---|---|---|---|
| A | 1 | 1 | 1 | 1 | 0 | 1 |
| B | 0 | 1 | 1 | 0 | 1 | 0 |
| C | 0 | 0 | 1 | 0 | 0 | 0 |
| D | 1 | 0 | 0 | 1 | 1 | 0 |
| E | 0 | 1 | 0 | 0 | 1 | 1 |
| F | 1 | 0 | 0 | 0 | 1 | 1 |

The information in the poset *OV* is completely represented by table 3 (Marchotorchino, 2002), this is not the case for poset *VC* witch is partially represented by table 4 and table 5. We don't know how to represent exactly its information in a single table. The theoretical answer can possibly be found in works on algebraic structures. Moreover, the relational representation of answer's sets is useful to define news similarity measures, as we will see.

## 5. Similarity measures for poset with relational relation

The similarity measures may be written with the relational notation, i.e. with adjacency matrix. It is the most efficient formulation to compute them easily. In the first part of this chapter we recall some of them and then we present a new measure, useful for posets comparisons.

### 5.1 Antichain's case

Let $S_1$ and $S_2$ be the two simple sets to be compared composed with elements $x_i$ ($i$ varying from 1 to n). They adjacency matrix are $\mathbf{A}_{S1}$ and $\mathbf{A}_{S2}$ (they are in fact vectors). The number of common element $|S_1 \cap S_2|$ is the product of the two matrix $\mathbf{A}_{S1}$ and $\mathbf{A}_{S2}$.

$$|S_1 \cap S_2| = \mathbf{A}_{S1}\mathbf{A}_{S2} = \sum_{i=1}^{n} a_{1i}.a_{2i} \tag{14}$$

### 5.2 Partition's Case

Let us V1 and V2 be the two partitions to be compared and there adjacency matrix be given respectively by $\mathbf{A}_{V1}$ and $\mathbf{A}_{V2}$. The $n \times n$ matrices $\mathbf{A}_{V1}$ and $\mathbf{A}_{V2}$ are defined as follows:

$$\mathbf{A}_{V1} = [a_{ij}] \quad \begin{cases} a_{ij} = 1 \text{ if } x_i \text{ and } x_j \text{ are in the same classe of } V1 \\ a_{ij} = 0 \text{ else} \end{cases}$$

$$\mathbf{A}_{V2} = [a_{ij}] \quad \begin{cases} a_{ij} = 1 \text{ if } x_i \text{ and } x_j \text{ are in the same classe of } V2 \\ a_{ij} = 0 \text{ else} \end{cases}$$

Let N be the corresponding contingency table of dimension pxq, N is composed with elements $n_{uv}$ ($u \in [1, p]$ and $v \in [1, q]$) representing the number of common terms in the class u of V1 and v of V2. We call $n_{u.}$ the number of items in the class u of V1 and $n_{.v}$ the number of items in the class v of V2.

With this notation, the Rand index will be written (Saporta and Youness, 2002) :

$$R_{Rand} = \frac{2A}{n(n-1)} = \frac{2\sum_{u=1}^{p}\sum_{v=1}^{q} n_{uv}^2 - \sum_{u=1}^{p} n_{u.}^2 - \sum_{v=1}^{q} n_{.v}^2 + n^2}{n^2} \tag{15}$$

Saporta et al. (Saporta and Youness, 2002) recall the linear formulas demonstrated by Kendall and Marcotorchino :

$$\sum_{u}\sum_{v} n_{uv}^2 = \sum_{i}\sum_{j} a_{ij}^1 a_{ij}^2 \tag{16}$$

$$\sum_{u} n_{u.}^2 = \sum_{i}\sum_{j} a_{ij}^1 \quad \text{and} \quad \sum_{v} n_{.v}^2 = \sum_{i}\sum_{j} a_{ij}^2 \tag{17}$$

So the Rand index can also be written:

$$R_{Rand} = \sum_{i=1}^{n} \sum_{j=1}^{n} \frac{\left(a_{ij}^1 a_{ij}^2 + \overline{a_{ij}^1 a_{ij}^2}\right)}{n^2} \text{ with } \overline{a} = 1 - a \quad (18)$$

In the same way the asymmetrical Rand index and the Jaccard index has been converted by Saporta in terms of relational notations and may be written:

$$R_{RandAsy}(V1,V2) = \frac{n^2 + \sum_{u=1}^{p} \sum_{v=1}^{q} n_{uv}^2 - \sum_{u=1}^{p} n_{u.}^2}{n^2} \quad (19)$$

or

$$R_{RandAsy}(V1,V2) = \frac{n^2 + \sum_i \sum_j a_{ij}^1 a_{ij}^2 - \sum_i \sum_j a_{ij}^1}{n^2} \quad (20)$$

$$J(V1,V2) = \frac{n + \sum_{u=1}^{p} \sum_{v=1}^{q} n_{uv}^2}{n + \sum_{u=1}^{p} \sum_{v=1}^{q} n_{uv}^2 - \sum_{u=1}^{p} n_{u.}^2 - \sum_{v=1}^{q} n_{.v}^2} \quad (21)$$

or with the relational notations:

$$J(V1,V2) = \frac{n + \sum_{i=1}^{n} \sum_{j=1}^{n} a_{ij}^1 a_{ij}^2}{n - \sum_{i=1}^{n} \sum_{j=1}^{n} \overline{a_{ij}^1 a_{ij}^2} - \sum_{i=1}^{n} \sum_{j=1}^{n} a_{ij}^1} \quad (22)$$

or

$$J(V1,V2) = \frac{n + \sum_{i=1}^{n} \sum_{j=1}^{n} a_{ij}^1 a_{ij}^2}{n - \sum_{i=1}^{n} \sum_{j=1}^{n} a_{ij}^1 \overline{a_{ij}^2} - \sum_{i=1}^{n} \sum_{j=1}^{n} a_{ij}^2} \quad (23)$$

### 5.3 Poset's case

#### 5.3.1 Ordered partition's case

Let $OV1 = (A1_u)_{u \in [1,p]}$ and $OV2 = (A2_v)_{v \in [1,q]}$ be the two ordered partitions to be compared. We will have :

$|A1_u \cap A2_v| = n_{uv}$

$|A1_u| = n_{u.}$ and $|A2_v| = n_{.v}$

So, by using the membership function $\varphi(i) = \frac{1}{2^{i-1}}$, the intersection, union and cardinality proposed in section 3.4 may be written:

$$|U_{OV1} \cap U_{OV2}| = \sum_{u=1}^{p} \sum_{v=1}^{q} \frac{n_{uv}}{2^{\max(u,v)-1}} \quad (24)$$

$$|U_{OV1}| = \sum_{u=1}^{p} \frac{n_{u.}}{2^{u-1}} \quad (25)$$

$$|U_{OV2}| = \sum_{v=1}^{q} \frac{n_{.v}}{2^{v-1}} \quad (26)$$

$$|U_{OV1} \cup U_{OV2}| = \sum_{u=1}^{p} \sum_{v=1}^{q} \frac{n_{uv}}{2^{u-1}} + \sum_{u=1}^{p} \sum_{v=1}^{j-1} \frac{n_{uv}}{2^{v-1}} + \sum_{u=1}^{p} \frac{n_{u.} - \sum_{v=1}^{q} n_{uv}}{2^{u-1}} + \sum_{v=1}^{q} \frac{n_{.v} - \sum_{u=1}^{p} n_{uv}}{2^{v-1}} \quad (27)$$

So it is possible to write the ordered similarity measures presented in (Egghe and Michel, 2003) by using theses new notations. We leave it to the interested reader.

*5.3.2 Case of partition of chain*

In the case of partition of chain we have not found any similarity measures useful for the evaluation of IR systems.

Let us remember that $OV_{VC}$ denote the poset constructed from the poset $VC$ with the relation $R_{rank}$. We call $V_{VC}$ the poset constructed with the relation $R_{\in}$ from the poset $VC$. $OV_{VC}$ is an ordered partition and $V_{VC}$ is a partition and so we can compute $J_F(OV_{VC}1, OV_{VC}2)$ and $R_{Rand}(V_{VC}1, V_{VC}2)$.

We assume that the similarity between two partition of chain VC1 and VC2 can be expressed by the product $J_F(OV_{VC}1, OV_{VC}2)$ and $R_{Rand}(V_{VC}1, V_{VC}2)$. So a similarity of partition of chain can be written as :

$$J(VC1, VC2) = R_{Rand}(V_{VC}1, V_{VC}2) \times J_F(OV_{VC}1, OV_{VC}2) \quad (28)$$

We will call this measure the similarity of Jaccard for poset. We assume that is possible to construct other similarity measures for poset from classical similarity measure, like the cosine, the recall, … etc., by using the same construction.

We make the following convention for the relational formula of $J(VC1, VC2)$ : let $u \in [1, p]$, $v \in [1, q]$, $w \in [1, r]$, $z \in [1, s]$, let $n^{V_{VC}}{}_{uv}$ be the number of common terms in the class u of $V_{VC}1$ and v of $V_{VC}2$, let $n^{V_{VC}}{}_{u.}$ be the number of items in the class u of $V_{VC}1$ and let $n^{V_{VC}}{}_{.v}$ be the number of items in the class v of $V_{VC}2$, let $n^{OV_{VC}}{}_{uv}$ be the number of common terms in the class w of $OV_{VC}1$ and z of $OV_{VC}2$, let $n^{OV_{VC}}{}_{w.}$ be the number of items in the class w of $OV_{VC}1$ and let $n^{OV_{VC}}{}_{.z}$ be the number of items in the class z of $OV_{VC}2$.

We have :

$$R_{Rand}(V_{VC}1, V_{VC}2) = \frac{2\sum_{u=1}^{p}\sum_{v=1}^{q} n^{V_{vc}}{}_{uv}{}^2 - \sum_{u=1}^{p} n^{V_{vc}}{}_{u.}{}^2 - \sum_{v=1}^{q} n^{V_{vc}}{}_{.v}{}^2 + n^2}{n^2} \quad (29)$$

$$J_F(OV_{VC}1, OV_{VC}2) = \frac{\sum_{w=1}^{r}\sum_{z=1}^{s} \frac{n_{wz}}{2^{\max(w,z)-1}}}{\alpha} \quad (30)$$

with
$$\alpha = \sum_{w=1}^{r}\sum_{z=1}^{s}\frac{n^{OV_{vc}}{}_{wz}}{2^{w-1}} + \sum_{w=1}^{r}\sum_{z=1}^{z-1}\frac{n^{OV_{vc}}{}_{wz}}{2^{z-1}} + \sum_{u=1}^{r}\frac{n^{OV_{vc}}{}_{w.} - \sum_{z=1}^{s} n^{OV_{vc}}{}_{wz}}{2^{w-1}} + \sum_{v=1}^{s}\frac{n^{OV_{vc}}{}_{.z} - \sum_{w=1}^{r} n^{OV_{vc}}{}_{wz}}{2^{z-1}} \quad (31)$$

which gives the relational notation for $J(VC1, VC2)$.

## 6. Conclusion

By applying the poset representation in terms of graph and numerical data we have presented a method to construct similarity measures useful in the case of IR systems evaluation and valid for partitions with partial orders. We are sure that the measures presented in the case of ordered partition poset are valid because we have only modified the notation. We expect that the measures proposed in the case of partition of chains are valid too but we have to verify it in a theoretical and experimental way. In future work, we plan to compare the similarity defined in this paper with the similarity measures defined between graph and relations in order to determine how they can be used in the case of IR systems evaluation. Finally, we will investigate whether the representation by weighted graph (Umeyama, 1988) is efficient for our problem.

## 7. Acknowledgments

I want to thanks P. Kantor, T. Saracevic and N. Belkin form the Rutgers Distributed Lab. For Digital Library (SCILS, Rutgers University, NJ, USA) for their hospitality, the excellent working conditions and their useful comments regarding this paper and JF Marcotorchino, (Thales, France) for having introduced me to the relational process of computation.